\documentclass[aps,pra,reprint,showpacs,amsmath,amssymb]{revtex4-1}

\usepackage{graphicx}
\usepackage{bm}
\usepackage{amssymb}
\usepackage{amsmath}
\usepackage{amsthm}
\usepackage{amsfonts}
\usepackage{mathrsfs}
\usepackage{hyperref}

\begin{document}

\title{Wavelength attack on practical continuous-variable quantum-key-distribution system with a heterodyne protocol}

\author{Xiang-Chun Ma, Shi-Hai Sun, Mu-Sheng Jiang}
\author{Lin-Mei Liang}\email{nmliang@nudt.edu.cn}
\affiliation{Department of Physics, National University of Defense Technology, Changsha 410073, People's Republic of China}

\date{\today}

\begin{abstract}
We present the wavelength attack on a practical continuous-variable quantum-key-distribution system with a heterodyne protocol, in which the transmittance of beam splitters at Bob's station is wavelength-dependent. Our strategy is proposed independent of but analogous to that of Huang \textit{et al}. [arXiv: 1206.6550v1 [quant-ph]], but in that paper the shot noise of the two beams that Eve sends to Bob, transmitting after the homodyne detector, is unconsidered. However, shot noise is the main contribution to the deviation of Bob's measurements from Eve's when implementing the wavelength attack, so it must be considered accurately. In this paper, we firstly analyze the solutions of the equations specifically that must be satisfied in this attack, which is not considered rigorously by Huang \textit{et al}. Then we calculate the shot noise of the homodyne detector accurately and conclude that the wavelength attack can be implemented successfully in some parameter regime.
\end{abstract}
\pacs{03.67.Hk, 03.67.Dd, 42.50.--p, 89.70.Cf} 

\maketitle

\section{\label{sec:Intro}Introduction}
Continuous-variable quantum key distribution (CVQKD), which is an alternative to single-photon quantum-key distribution (QKD), has many advantages, such as high repetition rate of communication, high detection efficiency, and ease of integration with standard telecom components, so it has received much more attention in recent years \cite{Ral99,Gro03,Lod07,Lev09,She10,Jou11,Wee12}. However, because there exist some imperfections such as loss or noise in practical system, actually the unconditional security of the final key of QKD may be compromised. It has been extensively investigated in single-photon QKD, such as photon-number-splitting (PNS) attack, passive Faraday-mirror attack and parially random phase attack etc.~\cite{Mak06,Zha08,Sun11,Liu11,Sun12}, but not the case in CVQKD \cite{Lod07L,Sun12A}. This is because the system of CVQKD is a one-way communication system and needs fewer optical elements than two-way system. Furthermore, most of intervention of Eve on the system of CVQKD can be detected by the parameter estimation of classical postprocessing of CVQKD.

In Ref.~\cite{Li11}, the wavelength-dependent property of a beam splitter (BS) was exploited by eavesdropper Eve to attack single-photon QKD successfully. Subsequently, Huang \textit{et al} extended this attack, the so-called wavelength attack, to the all-fiber system of CVQKD \cite{Hua12}. However, in that paper, two significant problems are not considered. First, the equation that must be satisfied in this attack, so-called \textit{attacking equation}, was not solved specifically in some permitted parameter regime, which may make this attack invalid. Second, the shot noise of the two beams that Eve sends to Bob, when transmitting through the homodyne detector, was neglected. However, shot noise is the main contribution to the deviation of Bob's measurements from Eve's when implementing the wavelength attack, so it must be considered accurately.

In this paper, we demonstrate and resolve these two problems, and then we improve the wavelength attack method against the all-fiber CVQKD system by tuning the attacking parameter's regime. Finally, we conclude that the wavelength attack will be implemented successfully in some parameter regime. The paper is organized as follows: In Sec.~\ref{sec:Wave}, we demonstrate the wavelength attack and solve the attacking equations specifically. Then, we analyze the shot noise introduced by the one-port and two-port homodyne detectors and calculate the conditional variance between the two legitimate parties Alice and Bob considering the deviations introduced by the shot noise. Finally, in Sec.~\ref{sec:Discussion}, we discuss and make some conclusions about the feasibility of this wavelength attack on a practical CVQKD system based on the conditional variance obtained in Sec.~\ref{sec:Deviation}.

\section{\label{sec:Wave}Wavelength attack on practical CVQKD system}
\subsection{\label{sec:Scheme}Scheme of wavelength attack}
In a practical CVQKD system, Alice first modulates a coherent state $(\hat{x}_A,\hat{p}_A)$ by amplitude and phase modulators according to bivariate Gaussian distributions centered on $(x_A,p_A)$ of variance $V_AN_0$, where $N_0$ is the shot-noise variance that appears in the Heisenberg relation $\Delta x\Delta p\geq N_0$ \cite{Lod07},
\begin{equation}\label{eq:xApA}
 \hat{x}_A=x_A+\hat{x}_N^A,~
 \hat{p}_A=p_A+\hat{p}_N^A.
\end{equation}
$\langle(\hat{x}_N^A)^2\rangle=N_0$, $\langle(\hat{p}_N^A)^2\rangle=N_0$. Subsequently, she sends this state to Bob through a quantum channel optical fiber and Bob implements a homodyne or heterodyne detection after receiving this state. So, after many repetitions, Alice and Bob can share trains of correlated data and then get the final key after the classical data postprocessing procedure. However, because the practical system may have some imperfections which will leave a loophole for the potential eavesdropper Eve, we must calibrate the system carefully and make efforts to eliminate all the existing loopholes.

\begin{figure}
 \scalebox{1}{\includegraphics[width=\columnwidth]{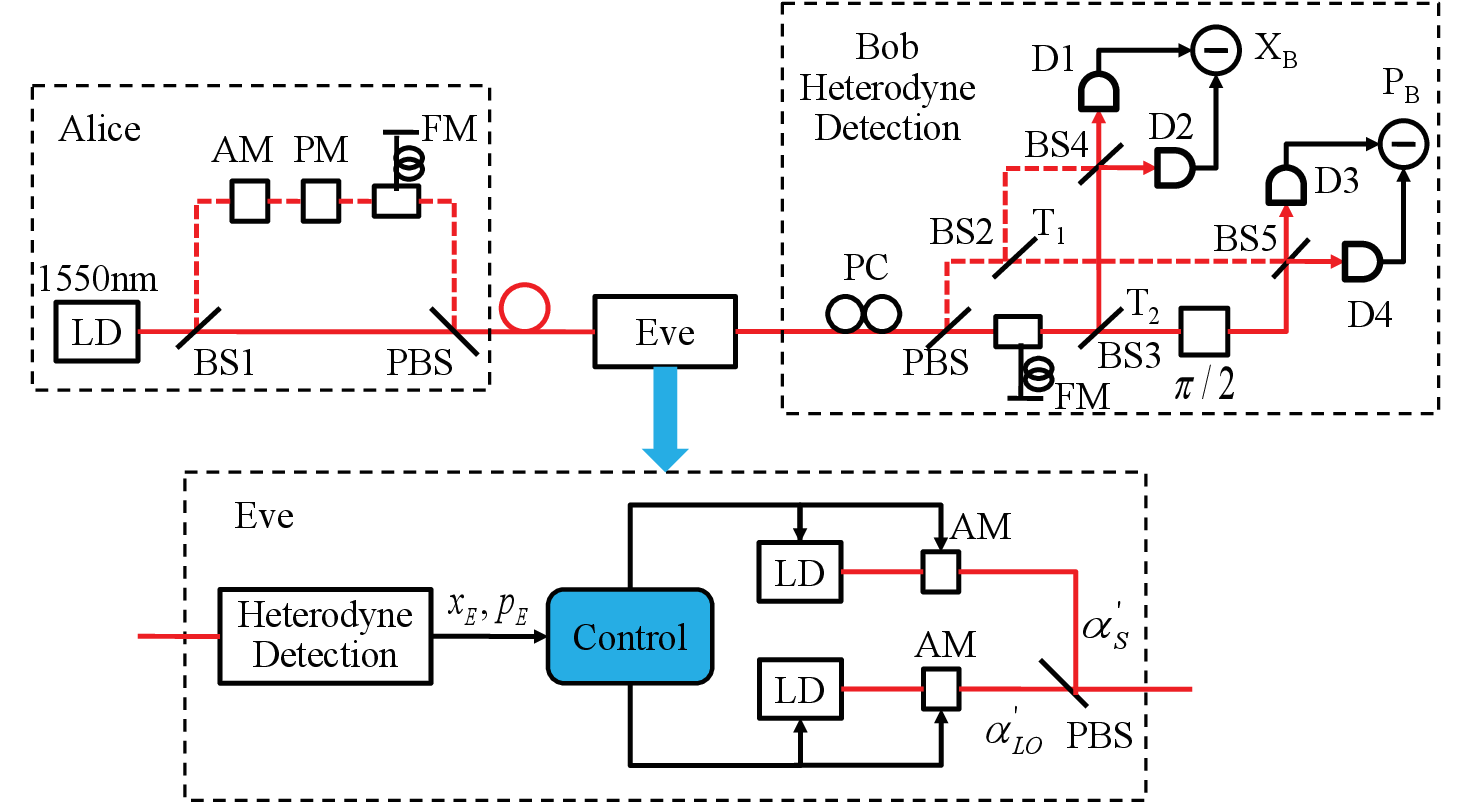}}
 \caption{\label{fig:1}(Color online) Wavelength attack setup on a practical CVQKD system with heterodyne detection (dashed line indicates signal beam and solid line denotes the local oscillator) \cite{Hua12}. LD, laser diode; AM: amplitude modulator; PM, phase modulator; FM, Faraday mirror; PBS, polarization beam splitter; PC, polarization controller; D, photodetector; and BS1-5, beam splitters. BS2-5 have the same wavelength-dependent property, i.e., when the signal beam (local oscillator) transmits through them, all of their respective transmittance are $T_1(T_2)$. BS1 is a beam splitter at Alice's side and is not necessarily wavelength dependent.}
\end{figure}

Generally, the transmittance of some practical beam splitters depends on the wavelength of beams, namely $T=F^2\sin^2(\frac{cw\lambda^{2.5}}{F})\equiv T(\lambda)$, where $F^2$ is the maximal power that is coupled, $c$ is the coupling coefficient, and $\omega$ is the heat source width \cite{Ank86,Tek90}. Consequently, proposed in Ref.~\cite{Hua12}, Eve can use this wavelength-dependent property to attack the practical CVQKD system with heterodyne protocol shown in Fig.~\ref{fig:1}. She can send two beams whose wavelength and intensity can be tuned to control Bob's measurement results, which are made identical to Eve's.

Specifically speaking, Eve first intercepts Alice's sending states and makes a heterodyne detection on them, so she can get the quadratures $x_E , p_E$, which can be given by
\begin{equation}\label{eq:xEpE}
 \hat{x}_E=\hat{x}_A+\hat{x}_N^E,~
 \hat{p}_E=\hat{p}_A+\hat{p}_N^E,
\end{equation}
where $\langle(\hat{x}_N^E)^2\rangle=N_0$, $\langle(\hat{p}_N^E)^2\rangle=N_0$ is the shot noise introduced by Eve's heterodyne detection. Then, she sends to Bob two beams whose intensities are denoted as $|\alpha'_S|^2$ and $|\alpha'_{L\!O}|^2$ respectively. Because the interference between these two beams is destroyed, Eve might be able to control the intensities and wavelengths of them to make these two beams, after transmitting Bob's heterodyne detectors, satisfy
\begin{equation}\label{eq:attackeq}
\begin{split}
&(1\negmedspace-\negmedspace T_1)\negmedspace(1\negmedspace-\negmedspace2T_1)\negmedspace|\alpha'_S|^2\negmedspace-\negmedspace(1\negmedspace-\negmedspace T_2)\negmedspace(1\negmedspace-\negmedspace2T_2)\negmedspace|\alpha'_{L\!O}|^2\negmedspace=\negmedspace\sqrt{\eta}x_E|\alpha_{L\!O}|,\\
&T_1(1-2T_1)|\alpha'_S|^2-T_2(1-2T_2)|\alpha'_{L\!O}|^2=\sqrt{\eta}p_E|\alpha_{L\!O}|,
\end{split}
\end{equation}
where $T_1(T_2)$ represents the transmittance of a practical beam splitter corresponding to the beam $\alpha'_S(\alpha'_{L\!O})$ whose wavelength is $\lambda_1(\lambda_2)$. $\alpha_{L\!O}$ is the amplitude of local oscillator in the absence of this wavelength attack, and $\eta$ is the channel loss. After scaling with $\sqrt{2}\alpha_{L\!O}$ \cite{Hua12}, then, Bob will get measurements $\sqrt{\eta}x_E/\sqrt{2}$ and $\sqrt{\eta}p_E/\sqrt{2}$. Thus, the wavelength attack may succeed; however, whether Eqs.~(\ref{eq:attackeq}) have the real solutions in the practical parameter regime determines the validity of this wavelength attack. So, in what follows we analytically investigate the solutions of Eqs.~(\ref{eq:attackeq}) in the permitted parameter regime.

First, we rewrite Eqs.~(\ref{eq:attackeq}) as
\begin{equation}\label{eq:ivattackeq}
\begin{split}
&(1\negmedspace-\negmedspace T_1)\negmedspace(1\negmedspace-\negmedspace2T_1)\negmedspace|\alpha'_S|^2\negmedspace=\negmedspace\sqrt{\eta}x_E|\alpha_{L\!O}|\negmedspace+\negmedspace(1\negmedspace-\negmedspace T_2)\negmedspace(1\negmedspace-\negmedspace2T_2)\negmedspace|\alpha'_{L\!O}|^2,\\
&T_1(1-2T_1)|\alpha'_S|^2=\sqrt{\eta}p_E|\alpha_{L\!O}|+T_2(1-2T_2)|\alpha'_{L\!O}|^2.
\end{split}
\end{equation}
Provided that $\alpha'_{L\!O}$ is the same as $\alpha_{L\!O}$, then $T_2$ is $1/2$ and Eqs.~(\ref{eq:ivattackeq}) will be reduced to
\begin{equation}\label{eq:redateq}
\begin{split}
(1-T_1)(1-2T_1)|\alpha'_S|^2=\sqrt{\eta}x_E|\alpha_{L\!O}|,\\
T_1(1-2T_1)|\alpha'_S|^2=\sqrt{\eta}p_E|\alpha_{L\!O}|.
\end{split}
\end{equation}
As Ref.~\cite{Hua12} says, generally $x_E , p_E$ are very small in practical implementation of CVQKD, and Eqs.~(\ref{eq:redateq}) are always solvable if $x_E , p_E$ are both positive or negative at the same time. That is,
\begin{equation}\label{eq:T1solu}
\frac{1-T_1}{T_1}=\frac{x_E}{p_E},~T_1=\frac{p_E}{x_E+p_E}\in[0,1].
\end{equation}
However, when $x_E , p_E$ are different in sign, $T_1\overline{\in}[0,1]$, but by virtue of opting for an appropriate $T_2$ ($\neq1/2$), we are always able to confirm both the right-hand sides of Eqs.~(\ref{eq:ivattackeq}) as being either positive or negative at the same time, because both the second terms on the right-hand sides of Eqs.~(\ref{eq:ivattackeq}) are the same in sign. Hence, Eqs.~(\ref{eq:ivattackeq}) or Eqs.~(\ref{eq:attackeq}) always hold if we select appropriate $\alpha'_{L\!O}$.

Additionally, we point out that we can always make the-right hand side of Eqs.~(\ref{eq:ivattackeq}) sufficiently small and $T_2$ close to $1/2$; thus $|\alpha'_S|^2$ in the left-hand of Eqs.~(\ref{eq:ivattackeq}) can always be small too, especially in discrete modulation protocol of CVQKD \cite{Lev09,She10} in which the signal intensity is always on a single-photon level. Consequently, this attack cannot be avoided efficiently even if Bob added a wavelength filter on his system before detectors, which is the same case in Ref.~\cite{Li11}. This is because such extremely weak signal is still able to transmit through the practical filter just by increasing the intensity of the incoming light, and the wavelength of the fake local oscillator can be close to the original one which is 1550 nm. The laser after transmitting through the practical filter is permitted to have a line width, so the fake local oscillator cannot be filtered either.

Consequently, by implementing this wavelength attack, theoretically Eve can control her attacking parameters to make Bob's measurements identical to hers, or rather Eve completely knows Bob's measurements. However, in Eqs.~(\ref{eq:attackeq}) we do not consider the interference between Eve's sending beams and vacuum mode entering from other ports of Bob's beam splitters, as Fig.~\ref{fig:1} shows. This interference could introduce excess noise into Bob's measurements, thus leading them to deviate from Eve's. We demonstrate this in the next section.

\subsection{\label{sec:Deviation}Deviation of Bob's measurements from Eve's and Alice's}
As the previous section demonstrated, the interference with vacuum mode will lead Bob's measurements to deviate from Eve's. If the deviation is unfortunately large, Alice and Bob will find that they cannot distill any secure keys because Bob's measurements are too noisy after comparing partial data within parameter-estimation phase of data postprocessing. So, we must calculate the conditional variance between Alice and Bob to see whether Eve's wavelength attack could be successful. We point out that the shot noise introduced by Bob's detectors is the main contribution to the conditional variance, which is not noted in Ref.~\cite{Hua12}. We begin this calculation by analyzing the quantum noise of unbalanced homodyne detectors first and apply the results to Bob's apparatus.

\subsubsection{\label{sec:Quantum}Quantum noise on unbalanced homodyne detector}
\begin{figure}[h]
 \scalebox{1}{\includegraphics[width=\columnwidth]{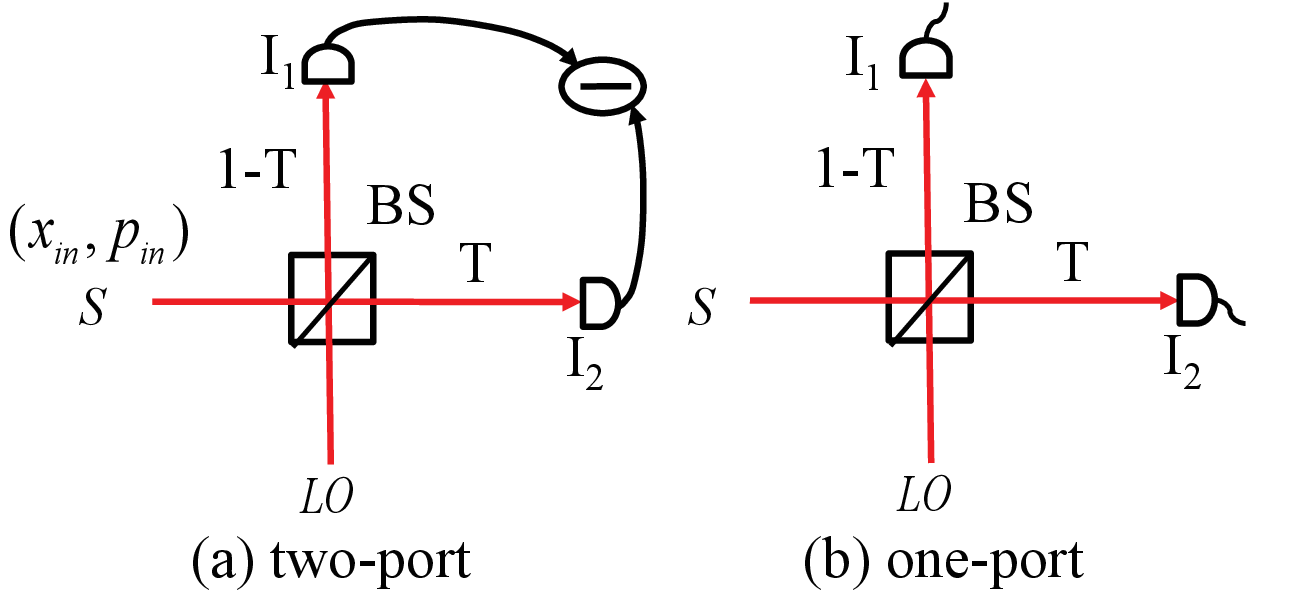}}
 \caption{\label{fig:2}(Color online) Unbalanced homodyne detector with (a) two-port or (b) one-port.}
\end{figure}

Pulsed homodyne detector is extensively exploited to measure a weak signal with a bright local oscillator \cite{Yue83,Sch84,Ray95}. As depicted in Fig.~\ref{fig:2}, when the transmittance of beam splitter $T\!=\!1/2$, the homodyne detector is balanced homodyne detector (BHD), or unbalanced homodyne detector (UBHD). Figure \ref{fig:2}(a) shows two-port homodyne detector with a subtractor, and Fig.~\ref{fig:2}(b) shows one-port homodyne detector without a subtractor. Those two-port or one-port homodyne detector can be used to measure a weak signal's quadratures or vacuum state's quantum noise.

Generally, two-port balanced homodyne detector can measure the quadratures $x_A$ or $p_A$ of weak signal $\alpha_S$, and between $\alpha_S$ and the local oscillator the relative phase is $\theta$. The measurement is
\begin{equation}\label{eq:x0}
x_\theta=2|\alpha_{L\!O}|(x_{in}\cos\theta+p_{in}\sin\theta),
\end{equation}
and $x_{in}=x_A+x_N$, $p_{in}=p_A+p_N$. $x_N$, $p_N$ are the vacuum mode quadratures, of which the variance is $N_0$ \cite{Ray95}. The variance of $x_{in},p_{in}$ is
\begin{equation}\label{eq:xin}
\begin{split}
&\langle x^2_{in}\rangle=\langle x^2_A\rangle+\langle x^2_N\rangle=V_A+N_0,\\
&\langle p^2_{in}\rangle=\langle p^2_A\rangle+\langle p^2_N\rangle=V_A+N_0,
\end{split}
\end{equation}
where $V_A$ is the signal's variance. When $V_A=0$, the output is the shot noise $N_0$ \cite{Shot}. However, with unbalanced two-port homodyne detector, the measurement will be different from the one in Eq.~(\ref{eq:x0}). The following is the analysis of this case.

Since the strong local oscillator can be treated as a classical field \cite{Yue83,Sch84,Ray95}, the amplitude of it can be denoted as
\begin{equation}\label{eq:signal}
\alpha_{L\!O}=|\alpha_{L\!O}|e^{i\theta},
\end{equation}
where $\theta$ is the relative phase in Eq.~(\ref{eq:x0}). Provided the pulsed local oscillator $\alpha_{L\!O}$ and signal beam $\alpha_S$ have the same optical frequency and are all in the coherent state respectively, they will interfere with each other when transmitting through the beam splitter [cf. Fig.~\ref{fig:2}(a)]; i.e., after optical mixing, their intensity can be written as
\begin{widetext}
\begin{equation}\label{eq:I11}
\begin{split}
I_1&=|\sqrt{1-T}\alpha_S+\sqrt{T}\alpha_{L\!O}|^2=(1-T)|\alpha_S|^2+T|\alpha_{L\!O}|^2
+\sqrt{T(1-T)}(\alpha_S^*\alpha_{L\!O}+\alpha_{L\!O}^*\alpha_S)\\
&=(1-T)|\alpha_S|^2+T|\alpha_{L\!O}|^2+\sqrt{T(1-T)}\times2|\alpha_{L\!O}|(x_{in}\cos\theta+p_{in}\sin\theta),\\
I_2&=|\sqrt{T}\alpha_S-\sqrt{1-T}\alpha_{L\!O}|^2=T|\alpha_S|^2+(1-T)|\alpha_{L\!O}|^2
-\sqrt{T(1-T)}(\alpha_S^*\alpha_{L\!O}+\alpha_{L\!O}^*\alpha_S)\\
&=T|\alpha_S|^2+(1-T)|\alpha_{L\!O}|^2-\sqrt{T(1-T)}\times2|\alpha_{L\!O}|(x_{in}\cos\theta+p_{in}\sin\theta),
\end{split}
\end{equation}
\end{widetext}
where we have denoted $\frac{\alpha_S+\alpha^*_S}{2},\frac{\alpha_S-\alpha^*_S}{2i}$ by $x_{in}, p_{in}$, respectively. Then, with the subtraction of $I_1$ and $I_2$, the output of two-port UBHD can be obtained as
\begin{equation}\label{eq:X01}
X_\theta=2\sqrt{T(1-T)}x_\theta+(1-2T)(|\alpha_S|^2-|\alpha_{L\!O}|^2),
\end{equation}
where $x_\theta$ has been given by Eq.~(\ref{eq:x0}) and when $T=1/2$ Eq.~(\ref{eq:X01}) is reduced to Eq.~(\ref{eq:x0}). If the weak signal is a vacuum state, the amplitude of it can be read as $\alpha_S\!=\!\left<\alpha_S\right>\!+\!\delta\alpha_S$, and $\left<\alpha_S\right>\!=\!0$, so $x_A=p_A=0$ and $|\alpha_S|^2\!=\!|\delta\alpha_S|^2\equiv|\delta\alpha|^2<<|\alpha_{L\!O}|^2$. $\delta\alpha$ describes the amplitude fluctuation of the vacuum state. Thus, we can neglect its square terms, so the output of UBHD is
\begin{equation}\label{eq:X02}
X_\theta=2\sqrt{T(1-T)}x'_\theta-(1-2T)|\alpha_{L\!O}|^2,
\end{equation}
where $x'_\theta$ is obtained from Eq.~(\ref{eq:x0}) by $x_A=p_A=0$. When $\theta$ is selected to be $0$ or $\pi/2$, we can get the shot noise of the output $4T(1-T)N_0$ denoted as $(\Delta X)^2$ or $(\Delta P)^2$ [$(\Delta X)^2=(\Delta P)^2=4T(1-T)N_0$], which is consistent with the one of BHD output when $T=1/2$. The second term of the right-hand of Eq.~(\ref{eq:X02}) is the two ports' subtraction of the intensity of the local oscillator because of the unbalanced splitting rate of the asymmetric beam splitter (ABS).

Subsequently, let us analyze the unbalanced one-port homodyne detector. As shown in Fig.~\ref{fig:2}(b), the intensity of local oscillator or signal after transmitting the unbalanced beam splitter has been obtained already by Eq.~(\ref{eq:I11}). When the weak signal is a vacuum state, the intensity of Eq.~(\ref{eq:I11}) can be reduced to
\begin{equation}\label{eq:I12}
\begin{split}
I_1&=T|\alpha_{L\!O}|^2+2\sqrt{T(1-T)}|\alpha_{L\!O}|X_N,\\
I_2&=(1-T)|\alpha_{L\!O}|^2-2\sqrt{T(1-T)}|\alpha_{L\!O}|X_N,\\
\end{split}
\end{equation}
where $X_N\in\{x_N,p_N\}$ and $\theta$ has been selected to be $0$ or $\pi/2$. The right-hand side of each of Eqs.~(\ref{eq:I12}), except for the first term, which is the unbalanced part of splitting of the bright local oscillator because of the asymmetric splitting rate of ABS, is the fluctuation of each port of unbalanced one-port homodyne detector respectively. With the two-port or one-port UBHD, we can compute the noise of the beams Eve sends to Bob in the next section. However, keep in mind that the noise of UBHD is introduced by the vacuum state from the other input port of the asymmetric beam splitter of UBHD.

\subsubsection{\label{sec:Convariance}Conditional variance between Alice and Bob}
By looking back at Fig.~\ref{fig:1} again, it is clear that there are two types of UBHD at Bob's station. On one hand, the first kind of ABS (BS2 and BS3), through which the signal or local oscillator transmits alone, can be viewed as a one-port UBHD, and each port output is the interference between the vacuum state and either signal or local oscillator. So, as analyzed in Sec.~\ref{sec:Quantum}, the intensity of the first kind of ABS output is (taking the signal beam as an example and the local oscillator as analogous),
\begin{equation}\label{eq:Ist}
\begin{split}
I_S^r&=(1-T_1)I_S-2\sqrt{T_1(1-T_1)I_S}X_N~,\\
I_S^t&=T_1I_S+2\sqrt{T_1(1-T_1)I_S}X_N~.
\end{split}
\end{equation}
$I_S$ is the intensity of the signal beam Eve sends ($I_S=|\alpha'_S|^2$). On the other hand, the second kind of ABS (BS4 and BS5) is a two-port UBHD, but the output is the interference between signal beam and vacuum state in addition to the one between local oscillator and vacuum state, because the wavelengths of signal beam and local oscillator are different from each other and they cannot interfere with each other. Recall that the shot-noise amplitude of the two-port UBHD output is $\Delta\!X_T$ or $\Delta\!P_T$ as shown in Eq.~(\ref{eq:X02}). Scaled with $\sqrt{2}|\alpha_{L\!O}|$ [cf. Eq.~(\ref{eq:x0}), generally, for heterodyne detection local oscillator is split into two beams so its intensity should be divided by two], the measurements of Bob's detection are
\begin{widetext}
\begin{equation}\label{eq:xBpB}
\begin{split}
\hat{x}_B&=\frac{\Delta i}{\sqrt{2}q|\alpha_{L\!O}|}=\frac{(1-2T_1)I_S^r-(1-2T_2)I_{L\!O}^r+2\sqrt{I_S^r}\Delta\!X_S+2\sqrt{I_{L\!O}^r}\Delta\!X_{L\!O}}{\sqrt{2}|\alpha_{L\!O}|}\\
&=\sqrt{\frac{\eta}{2}}\hat{x}_E+\frac{(1-2T_1)[-2\sqrt{T_1(1-T_1)I_S}X_N]-(1-2T_2)[-2\sqrt{T_2(1-T_2)I_{L\!O}}X_N]+2\sqrt{I_S^r}\Delta\! X_S+2\sqrt{I_{L\!O}^r}\Delta\!X_{L\!O}}{\sqrt{2}|\alpha_{L\!O}|}\\
&=\sqrt{\frac{\eta}{2}}\hat{x}_E+\hat{x}_{B|E}~,\\
\hat{p}_B&=\frac{\Delta i}{\sqrt{2}q|\alpha_{L\!O}|}=\frac{(1-2T_1)I_S^t-(1-2T_2)I_{L\!O}^t+2\sqrt{I_S^t}\Delta\! P_S+2\sqrt{I_{L\!O}^t}\Delta\!P_{L\!O}}{\sqrt{2}|\alpha_{L\!O}|}\\
&=\sqrt{\frac{\eta}{2}}\hat{p}_E+\frac{(1\!-\!2T_1)\![2\sqrt{T_1(1\!-\!T_1)I_S}X_N]\!-\!(1\!-\!2T_2)\![2\sqrt{T_2(1\!-\!T_2)I_{L\!O}}X_N]+2\sqrt{I_S^t}\Delta\!P_S+2\sqrt{I_{L\!O}^t}\Delta\!P_{L\!O}}{\sqrt{2}|\alpha_{L\!O}|}\\
&=\sqrt{\frac{\eta}{2}}\hat{p}_E+\hat{p}_{B|E}~,
\end{split}
\end{equation}
\end{widetext}
where $\Delta i$ is the photocurrent subtraction of two port outputs of UBHD proportional to the intensity difference of two beams and the proportional constant is \emph{q}. $\hat{x}_{B|E}$ or $\hat{p}_{B|E}$ is the deviation of $\hat{x}_B$ or $\hat{p}_B$ from $\hat{x}_E$ or $\hat{p}_E$. The conditional variance of Bob's measurements conditioned on Eve's can be computed as
\begin{widetext}
\begin{equation}\label{eq:VBE}
\begin{split}
V^x_{B|E}&=\negmedspace\langle\hat{x}_{B|E}^2\rangle\negmedspace\\
&=\negmedspace\frac{\langle\{(1\negmedspace-\negmedspace2T_1)[-2\sqrt{T_1(1\negmedspace-\negmedspace T_1)I_S}X_N]\}^2\rangle
\negmedspace+\negmedspace\langle\{(1\negmedspace-\negmedspace2T_2)[-2\sqrt{T_2(1\negmedspace-\negmedspace T_2)I_{L\!O}}X_N]\}^2\rangle
\negmedspace+\negmedspace4I_S^r(\Delta\! X_S)^2\negmedspace+\negmedspace4I_{L\!O}^r(\Delta\!X_{L\!O})^2}{2|\alpha_{L\!O}|^2}\\
&\approx\frac{2T_2(1-T_2)(1-2T_2)^2I_{L\!O}N_0+2(1-T_2)I_{L\!O}[4T_2(1-T_2)N_0]}{|\alpha_{L\!O}|^2}\\
&\approx2T_2(1-T_2)(1-2T_2)^2N_0+8T_2(1-T_2)^2N_0~,\\
V^p_{B|E}&\approx V^x_{B|E}\equiv V_{B|E}~.
\end{split}
\end{equation}
\end{widetext}
The first approximate equality holds because $I_S,I_S^r<<|\alpha_{L\!O}|^2$, and the last equality of $V^x_{B|E}$ can be reduced for providing that the intensity of the fake local oscillator ($I_{L\!O}=|\alpha'_{L\!O}|^2$) that Eve sends is the same as that of the original local oscillator $(|\alpha_{L\!O}|^2)$. Under this assumption, $V^p_{B|E}$ is equivalent to $V^x_{B|E}$ and both of them can be denoted as $V_{B|E}$ for short.

In order to obtain the conditional variance ($V_{B|A}$) of Bob's measurements conditioned on Alice's random optional Gaussian variables $x_A$ or $p_A$, we can substitute Eqs.~(\ref{eq:xApA}) and (\ref{eq:xEpE}) into Eq.~(\ref{eq:xBpB}) to achieve Bob's measurements about Alice's mode. That's
\begin{equation}\label{eq:xBpB2}
\begin{split}
&\hat{x}_B=\sqrt{\frac{\eta}{2}}(x_A+\hat{x}_N^A+\hat{x}_N^E)+\hat{x}_{B|E},\\
&\hat{p}_B=\sqrt{\frac{\eta}{2}}(p_A+\hat{p}_N^A+\hat{p}_N^E)+\hat{p}_{B|E}.
\end{split}
\end{equation}
Then, we can get the conditional variance $V_{B|A}$
\begin{equation}\label{eq:VBA}
V_{B|A}=\langle(\hat{x}_B-\sqrt{\eta/2}x_A)^2\rangle=\eta N_0+V_{B|E}.
\end{equation}

Interestingly, it's clear that $V_{B|E}$ is always smaller than $V_{B|A}$, so the information between Bob and Eve is always much larger than that between Bob and Alice. This is consistent with the intercept-and-resend strategy; i.e., when implementing this wavelength attack, Eve totally knows Bob's measurements except for some small deviations. We analyze this deviation accurately and investigate how Eve hides herself with this attack in the next section.

\section{\label{sec:Discussion}Discussion and conclusion}
\begin{figure}[h]
 \scalebox{1}{\includegraphics[width=\columnwidth]{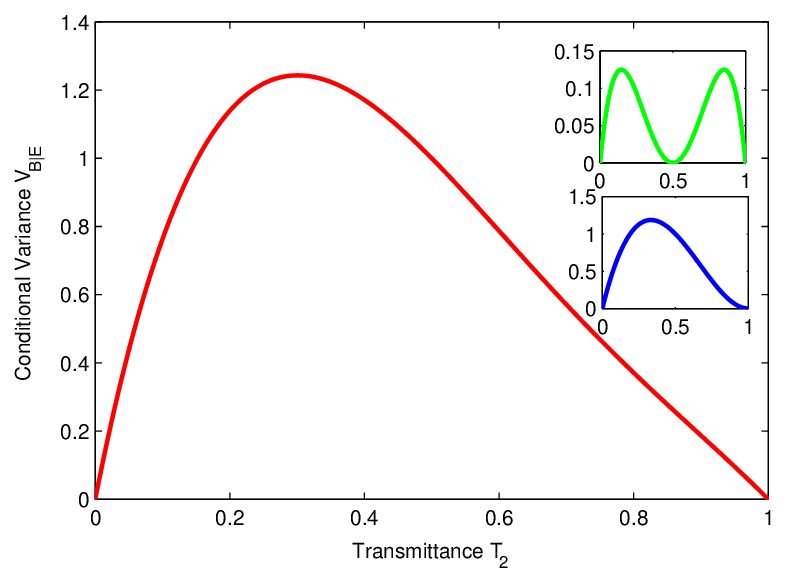}}
 \caption{\label{fig:3}(Color online) Conditional variance $V_{B|E}$ vs the transmittance $T_2$ of ABS on Bob's side. Inset shows the first term (top) and the second term (bottom) of $V_{B|E}$ in Eq.~(\ref{eq:VBE}).}
\end{figure}

As shown in previous sections, if Eve implements this wavelength attack against Bob's practical system, the conditional variance $V_{B|E}$ in Eq.~(\ref{eq:VBE}) versus the transmittance $T_2$ (corresponding to the local oscillator) of ABS on Bob's side is plotted in Fig.~\ref{fig:3}.

When $T_2$ equals 0.15 or 0.5, $V_{B|E}$ will be $N_0$ and can reach the maximum value $1.24N_0$ when $T_2$ equals 0.3. In practical heterodyne protocol of CVQKD, the secure conditional variance $V_{B|A}$ is always $N_0$ except for some small excess noise $\varepsilon N_0$. Consequently, Eve must select appropriate $T_2$, making $V_{B|E}$ equal $(1-\eta)N_0$, and then she can hide herself completely and get all information between Alice and Bob. Moreover, as discussed in Sec.~\ref{sec:Wave}, different values of $T_2$  can always make Eq.~(\ref{eq:attackeq}) satisfied; namely Eve could implement this wavelength attack successfully in any case. Besides, from Eq.~(\ref{eq:VBE}), it can be seen that the main contribution to conditional variance $V_{B|E}$ is the shot noise of two beams (especially local oscillator $|\alpha'_{L\!O}|^2$, $|\alpha'_S|^2<<|\alpha'_{L\!O}|^2$, so the shot noise contributed by weak signal $\alpha'_S$ can be neglected) when transmitting through the two-port UBHD. The noise introduced by the first kind of ABS of Bob's side (as one-port UBHD) (Fig.~\ref{fig:3} upper inset) is very small ($<0.15N_0$) and signal's contribution can be neglected because of the large denominator $|\alpha_{L\!O}|^2$ proportional to $10^8N_0$ \cite{Lod07}. Additionally, if Eve could reduce the intensity of two sending beams which are both smaller than the original local oscillator $|\alpha_{L\!O}|^2$, the conditional variance $V_{B|E}$ can be decreased to any extent because the shot noise, after being enlarged by the intensity of beam $\alpha'_S$ or $\alpha'_{L\!O}$, becomes small. For this purpose and not being found, Eve can exploit the wavelength-dependent property of additional monitoring ABS (splitting rate may be 1:99) to make the monitor recording the intensity of the beam $\alpha'_S$ or $\alpha'_{L\!O}$ be unchanged.

In conclusion, we investigate the feasibility of a wavelength attack combined with the intercept-resend method on practical CVQKD system with heterodyne protocol and conclude that this attack will be implemented successfully if we choose appropriate transmittance $T_2$ corresponding to the appropriate wavelength of the fake local oscillator. This attack can be implemented successfully due to several reasons. First, we analyzed the solution of equations that two sending beams by Eve satisfied, and it can be clearly seen that the solutions satisfying all conditions exist. Second, the main contribution of deviation of Bob's measurements from Eve's is due to the fake local oscillator's enlarging shot noise and can be reduced by selecting appropriate transmittance $T_2$ or decreasing the intensity of fake local oscillator. These two aspects are not considered in Ref.~\cite{Hua12}, which may make their scheme invalid in some parameter regimes. Last but not least, all of the practical beam splitters at Bob's station have the same wavelength-dependent property. However, if Bob inserts optical filters on his system before receiving the light, this attack cannot be avoided efficiently, as analyzed in Sec.~\ref{sec:Scheme}, which is considered to be impossible in Ref.~\cite{Hua12}. So, using high-quality filters and wavelength-independent beam splitters in practical CVQKD systems and carefully monitoring the local oscillator are very important to confirm the security of the heterodyne protocol of CVQKD.

Additionally, we point out that, for simplicity, in this paper we do not consider the imperfections of Bob's homodyne detector such as detection efficiency, electronic noise and other excess noise etc. However, when taking these imperfections into account, Eve only needs to slightly adjust her attacking parameters like $\eta$ and $T_2$ or fake local oscillator's intensity $|\alpha'_{L\!O}|^2$, which thus is able to hide her wavelength attack.

\begin{acknowledgments}
This work is supported by the National Natural Science Foundation of China, Grant No. 61072071. L.-M.L. is supported by the Program for New Century Excellent Talents. X.-C.M. and M.-S.J. acknowledge support from NUDT under Grant No. kxk130201.
\end{acknowledgments}

%


\begin{thebibliography}{99}
\bibitem{Shot} Here the shot noise variance $N_0$ (also called shot noise or vaccuum noise) indicates the variance of vaccum-state quadratures and has been normalized with LO intensity or power. In this paper we all scale $N_0=1/4$ as done in Ref.~\cite{Lod07}, and in other papers $N_0$ has been also scaled to $1$ or $1/2$ due to the different definitions between the quadratures ($\hat{x}$ or $\hat{p}$) and creation or annihilation operator ($\hat{a}^\dag$ or $\hat{a}$). However, when not scaled, the shot noise is surely proportional to the locol oscillator power.
\end{thebibliography}
\end{document}